\documentclass[12pt,BCOR2mm,DIV14]{scrartcl}

\newcommand{\cc}{\texttt}
\newcommand{\pkg}{\texttt}
\newcommand{\proglang}{\texttt}
\newcommand{\xmin}{x_{\min}}

\usepackage{graphicx}
\usepackage{booktabs}
\usepackage{amsmath}
\usepackage{microtype}
\usepackage[utf8]{inputenc} 
\usepackage{algorithm}
\usepackage{hyperref}
\usepackage[british]{babel}
\usepackage[round,authoryear,comma,sort&compress,numbers]{natbib}       
\graphicspath{{Figures/}}

\author{Colin S. Gillespie\\Newcastle University}
\title{Fitting heavy tailed distributions: the \pkg{poweRlaw} package}

\usepackage{Sweave}
\begin{document}
\maketitle
\begin{abstract}
Over the last few years, the power law distribution has been used as
  the data generating mechanism in many disparate fields. However, at times the
  techniques used to fit the power law distribution have been inappropriate.
  This paper describes the \pkg{poweRlaw} \proglang{R} package, which makes
  fitting power laws and other heavy-tailed distributions straightforward. This
  package contains R functions for fitting, comparing and visualising heavy
  tailed distributions. Overall, it provides a principled approach to power law 
  fitting. 
\end{abstract}

\section{Introduction}\label{S1}

The nineteenth century Italian economist, Vilfredo Pareto, observed that many
processes do not follow the Gaussian distribution. This observation lead to the
so-called 80/20 rule, that is:
\begin{quote}
  \textit{80\% of all effects results from 20\% of all causes.}
\end{quote}
This rule has been used to describe a wide variety of phenomena. For example,
20\% of employees of any business are responsible for 80\% of productive output
or 20\% of all people own 80\% of all wealth. By the middle of the twentieth
century, examples of these heavy tailed distributions had been used to describe
the number of papers published by scientists, sizes of cities and word frequency
(see \cite{Keller2005} for references).

In a similar vein, in 1999 two ground-breaking papers were published in
Science and Nature \citep{Barabasi1999,Albert1999}. In the first, the key result
was that the distribution of hyper-links in the World Wide Web seemed to follow
a power law distribution. Essentially, the connectivity of web-pages, $k$,
decreased with rate $k^{\alpha}$. This suggested a large connection variance,
with a small number of large, key nodes. The second paper presented a model that
could generate these networks and coined the phrase \textit{scale-free}. This
phrase implicitly linked these networks to the physics of phase transitions.

\begin{figure}[t] 
  \centering
  \includegraphics[width=\textwidth]{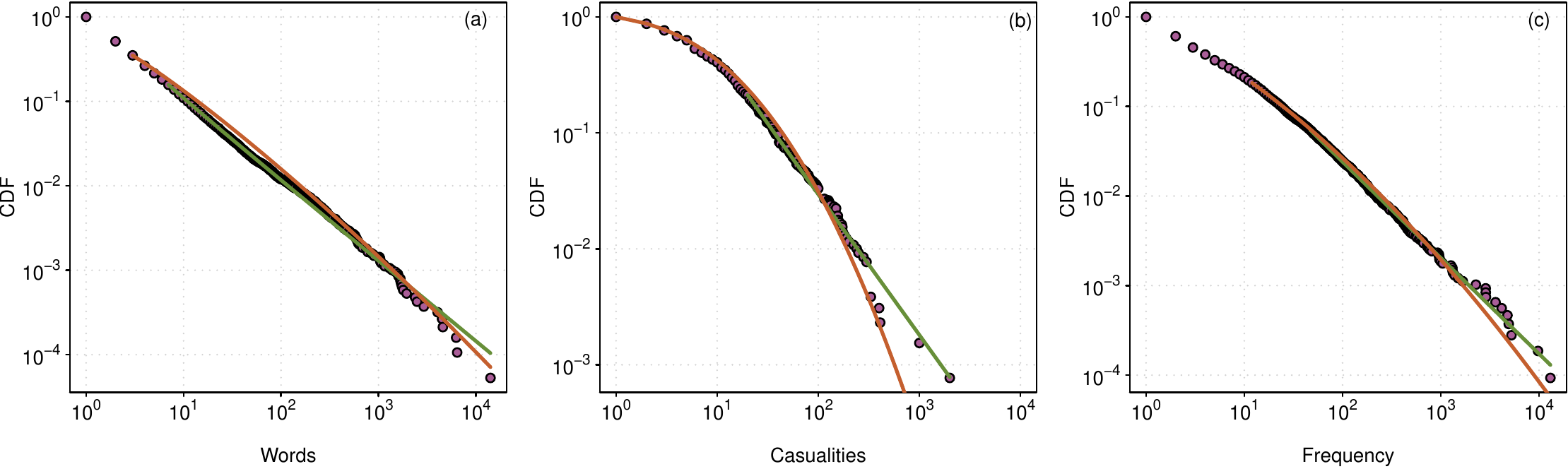}
  \caption{The cumulative distribution functions and their maximum likelihood
    power law (green) and log normal (orange) fit. Figure \textit{a}: Unique
    words in the novel Moby Dick. \textit{b}: Native American Casualities in the
    American Indian War. \textit{c}: Word frequency in the Swiss-Prot database
    (version 9). Further details of the data sets are given in
    Section~\ref{S1}.}\label{F1}
\end{figure}

Since these two landmark papers, there has been an explosion in supposed
scale-free phenomena (see Figure~\ref{F1}). For example,
\begin{itemize}
\item The occurrence of unique words in the novel \textit{Moby Dick} by Herman
  Melville \citep{Newman2005}.
\item Casualty numbers in armed conflicts \citep{Bohorquez2009, Friedman2013}.
\item Comparing manually curated databases with automatically curated biological
  databases \citep{Bell2012}.
\item Population sizes of cities \citep{Arcaute2013}.
\item Cliff rock fall scars \citep{Dewez2013}.
\item The number of interacting partners of proteins in yeast \citep{Yu2008}.
\item Movements of marine animals \citep{Sims2008,Edwards2012}.
\end{itemize}
Recently, this apparent ubiquity of power laws in a wide range of disparate
disciplines was questioned by \cite{Stumpf2012}. The authors point out that many
``observed'' power law relationships are highly suspect. In particular,
estimating the power law exponent on a log-log plot, whilst appealing, is a very
poor technique for fitting these types of models. Instead, a systematic,
principled and statistical rigorous approach should be applied (see
\cite{Goldstein2004}).

In this paper we describe the \pkg{poweRlaw} \proglang{R} package. This package
provides a straightforward interface to fitting power laws and other heavy
tailed distributions. Functions are provided for plotting, comparing
distributions and estimating parameter uncertainty.

\section{Mathematical background}

In this section, we introduce the discrete and continuous power law
distributions. We will discuss methods for fitting these distributions. While
this section just considers power law distributions, the techniques discussed
in Sections~\ref{2.2} and \ref{2.3} are general and can be applied to any
distribution.

\subsection{The power law distribution}

At the most basic level, there are two types of power law distribution: discrete
and continuous. The continuous version has probability density function (pdf)
\begin{equation}
p(x) = \frac{\alpha-1}{\xmin} \left(\frac{x}{\xmin}\right)^{-\alpha}
\end{equation}
where $\alpha >1$ and $\xmin >0$. While the discrete case has probability mass
function (pmf)
\begin{equation}
p(X=x) = \frac{x^{-\alpha}}{\zeta(\alpha, \xmin)}
\end{equation}
where
\begin{equation}
\zeta(\alpha, \xmin) = \sum_{n=0}^{\infty} (n + \xmin)^{-\alpha}
\end{equation}
is the generalised zeta function \citep{Abramowitz1970}.\footnote{The
  \pkg{poweRlaw} package uses the zeta function from the \pkg{VGAM} package to
  perform this calculation (see \cite{Yee2010}).} When $\xmin=1$,
$\zeta(\alpha, 1)$ is the standard zeta function. The cumulative density
functions have a relatively simple structure. For the continuous version we have
\begin{equation}
P(X\le x) = 1-\left(\frac{x}{\xmin}\right)^{-\alpha + 1} \;
\end{equation}
whilst for the discrete version we have
\begin{equation}
P(X \le x) = \frac{\zeta(\alpha, x)}{\zeta(\alpha, \xmin)} \;.
\end{equation}
The moments of the power law distribution are particularly interesting. For the
continuous power law we have
\[
E[X^m] = \int_{\xmin}^{\infty} x^m p(x)\, dx = \frac{\alpha - 1}{\alpha - 1 -m}
\xmin^m \;.
\]
So when 
\begin{itemize}
\item $1< \alpha \le 2$, all moments diverge, i.e. $E[X] = \infty$;
\item $2 < \alpha \le 3$, all second and higher-order moments diverge, i.e. $E[X^2] = \infty$;
\item $3 < \alpha \le m+1$, all $m$ and higher-order moments diverge, i.e. $E[X^m] = \infty$.
\end{itemize}

\subsection{Fitting heavy tailed distributions}\label{2.2}

To estimate the scaling the parameter $\alpha$ is relatively straightforward.
The maximum likelihood estimator (MLE) for the continuous power law is
\begin{equation}\label{6}
  \hat \alpha = 1 + n \left[\sum_{i=1}^n \ln \frac{x_i}{\xmin}\right]^{-1}
\end{equation}
where $x_i$ are the observed data values and $x_i \ge \xmin$
\citep{Muniruzzaman1957}. The discrete MLE of $\hat \alpha$ is not available,
instead we use the approximation
\begin{equation}\label{7}
\hat \alpha \simeq 1 + n \left[\sum_{i=1}^n \ln \frac{x_i}{\xmin - 0.5}\right]^{-1} \;.
\end{equation}
The discrete MLE approximation is identical to the exact continuous MLE, except
for the additional $0.5$ on the denominator.

When calculating the MLE for $\alpha$, we \textit{condition} on a particular
value of $\xmin$. When power laws are used in practice, it is usually argued
that only the tails of the distribution follow a power law, and so $\xmin$ must
be estimated. However as $\xmin$ increases, the amount data \textit{discarded}
also increases. So it clear that some care must be taken when choosing this
parameter.

The most common approach used to estimate $\xmin$ is from a visual inspection of
the data on a log-log plot. Clearly, this is highly subjective and error prone.
Instead, \cite{Clauset2009} recommend estimating the lower threshold using a
Kolmogorov-Smirnov approach. This statistic is simply the maximum distance
between the data and fitted model CDFs
\begin{equation}\label{8}
  D = \max_{x \ge \xmin} \vert S(x) - P(x) \vert
\end{equation}
where $S(x)$ and $P(x)$ are the CDFs of the data and model respectively (for $x
\ge \xmin$). The estimate of $\xmin$ is the value of $\xmin$ that minimises $D$.
This approach is completely general and can be used in conjunction with other
distributions.

\begin{algorithm}[t]
  \caption{Estimating the uncertainty in $\xmin$}\label{A1}
 \begin{tabular}{@{}ll@{}}
   {\small 1:} & Set $N$ equal to the number of values in the original data set \\
   {\small 2:} & \textbf{for} \texttt{i} in \texttt{1:B}:\\
   {\small 3:} & $\quad$ Sample $N$ values (with replacement) from the original data set \\
   {\small 4:} & $\quad$ Estimate $\xmin$ and $\alpha$ using the Kolmogorov-Smirnoff statistic\\
   {\small 5:} & \textbf{end for} \\
  \end{tabular}
\end{algorithm}

\subsection{Parameter uncertainty}\label{2.3}

For a particular value of $\xmin$, the standard error of the maximum likelihood
estimator for $\hat \alpha$ can be calculated analytically. However, to account
for the additional uncertainty of $\xmin$, it is necessary to use a
bootstrap procedure \citep{Efron1993}. Essentially, we sample with
replacement from the original data set and then re-infer the parameters at each
step (see Algorithm~\ref{A1}). The bootstrapping algorithm can be applied to any
distribution and can run in parallel.

\subsection{Alternative distributions}

The techniques discussed in the preceding sections provide flexible methods for
estimating distribution parameters and the lower cut-off, $\xmin$. In this
section, we discuss methods for testing whether the underlying distribution
could plausibly have a power law form.

Since it is possible to fit a power law distribution to \textit{any} data set,
it is appropriate to test whether the observed data actually follows a power
law. A standard goodness-of-fit test is to use bootstrapping to generate a
$p\,$-value to quantify the plausibility of the hypothesis. If the $p\,$-value is
large, than any difference between the empirical data and the model can be
explained with statistical fluctuations. If $p \simeq 0$, then the model does
not provide a plausible fit to the data and another distribution may be more
appropriate. When testing against the power law distribution the hypothesises
are:
\begin{align*}
&H_0: \text{data is generated from a power law distribution;}\\
&H_1: \text{data is not generated from a power law distribution.}
\end{align*}
The bootstrapping procedure is detailed in Algorithm~\ref{A2}. Essentially, we perform a
hypothesis test by generating multiple data sets (with parameters $\xmin$ and
$\alpha$) and then ``re-inferring" the model parameters. However, this technique
does have computational issues. In particular, when the scaling parameter
$\alpha \le 2$, the first moment (i.e. $E[X]$) is infinite and so extremely large
values frequently occur. Since generating random numbers for the discrete
power law distributions involves partitioning the cumulative density this may
make this approach unsuitable. 

\begin{algorithm}[t]
  \caption{Testing the power law hypothesis}\label{A2}
  \begin{tabular}{@{}ll@{}}
    {\small 1:} & Calculate point estimates for $\xmin$ and the scaling parameter $\alpha$ \\
    {\small 2:} & Calculate the Kolmogorov-Smirnov statistic, $KS_d$, for the original data set\\
    {\small 3:} & Set $n_1$ equal to the number of values below $\xmin$ \\
    {\small 4:} & Set $n_2 = n - n_1$ and $P = 0$\\
    {\small 5:} & \textbf{for} \texttt{i} in \texttt{1:B}:\\
    {\small 6:} & $\quad$ Simulate $n_1$ values from a uniform distribution:
    $U(1, \xmin)$ and $n_2$ values  \\
    &$\qquad$  from a  power law distribution (with parameter $\alpha$)\\
    {\small 7:} & $\quad$ Calculate the associated Kolmogorov-Smirnov statistic, $KS_{sim}$\\
    {\small 8:} & $\quad$ If $KS_d > KS_{sim}$, then $P = P + 1$\\
    {\small 9:} & \textbf{end for} \\
    {\small 10:} & $P= P/B$\\
  \end{tabular}
\end{algorithm}

An alternative  approach to assessing the power law model is a direct comparison with
another model. A standard technique is to use Vuong's test, which is a
likelihood ratio test for model selection using the Kullback-Leibler criteria.
The test statistic, $R$, is the ratio of the log likelihoods of the data between
the two competing models. The sign of $R$ indicates which model is
\textit{better}. Since the value of $R$ is subject to error, we use the method
proposed by \cite{Vuong1989}. See Appendix C in \cite{Clauset2009} for further
details.

\clearpage

\section{Example: word frequency in Moby Dick}

This example investigates the frequency of occurrence of unique words in the
novel Moby Dick by Herman Melville \citep{Clauset2009,Newman2005}. The data can
be downloaded from
\begin{center}
\url{http://tuvalu.santafe.edu/~aaronc/powerlaws/data.htm}
\end{center}
\noindent or directly loaded from the \pkg{poweRlaw} package
\begin{Schunk}
\begin{Sinput}
R> library("poweRlaw")
R> data("moby")
\end{Sinput}
\end{Schunk}
This data set contains the frequency of 18855 words. The most
commonly occurring word occurred 14086 times.

\subsection{Fitting a discrete power law}

To fit a discrete power law, we first create a discrete power law object using the \texttt{displ} constructor\footnote{\texttt{displ}: \textbf{dis}crete \textbf{p}ower \textbf{l}aw.}
\begin{Schunk}
\begin{Sinput}
R> pl_m = displ$new(moby)
\end{Sinput}
\end{Schunk}
The object \cc{pl\_m} is a S4 reference object. Initially the lower cut-off,
$\xmin$, is set to the smallest $x$ value and the scaling parameter, $\alpha$, is
set to \texttt{NULL}
\begin{Schunk}
\begin{Sinput}
R> pl_m$getXmin()
\end{Sinput}
\begin{Soutput}
[1] 1
\end{Soutput}
\begin{Sinput}
R> pl_m$getPars()
\end{Sinput}
\begin{Soutput}
NULL
\end{Soutput}
\end{Schunk}
\noindent The object also has standard setters
\begin{Schunk}
\begin{Sinput}
R> pl_m$setXmin(5)
R> pl_m$setPars(2)
\end{Sinput}
\end{Schunk}
\noindent For a given $\xmin$ value, we can estimate the corresponding $\alpha$
value using its maximum likelihood estimator
\begin{Schunk}
\begin{Sinput}
R> estimate_pars(pl_m)
\end{Sinput}
\begin{Soutput}
$pars
[1] 1.926

$value
[1] 14873

$counts
function gradient 
       5        5 

$convergence
[1] 0

$message
[1] "CONVERGENCE: REL_REDUCTION_OF_F <= FACTR*EPSMCH"

attr(,"class")
[1] "estimate_pars"
\end{Soutput}
\end{Schunk}
Alternatively, we can estimate the exponent using a parameter scan
\begin{Schunk}
\begin{Sinput}
R> estimate_pars(pl_m, pars=seq(1.5, 2.5, 0.01))
\end{Sinput}
\end{Schunk}
To estimate the lower bound $\xmin$, we use the Kolmogorov-Smirnoff approach
described in Section~\ref{2.2}
\begin{Schunk}
\begin{Sinput}
R> (est_pl = estimate_xmin(pl_m))
\end{Sinput}
\begin{Soutput}
$KS
[1] 0.008253

$xmin
[1] 7

$pars
[1] 1.953

attr(,"class")
[1] "estimate_xmin"
\end{Soutput}
\end{Schunk}
For the Moby Dick data set, the minimum is achieved when $\xmin=7$ and $D(7) =
0.00825$. Similar to the \texttt{estimate\_pars}
functions we can limit the search space using the \texttt{xmin} and
\texttt{pars} arguments.

To set the power law object to these optimal values, we just use the \cc{xmin} setter
\begin{Schunk}
\begin{Sinput}
R> pl_m$setXmin(est_pl)
\end{Sinput}
\end{Schunk}
To allow the user to explore different distributions and model fits, all
distribution objects have generic plot methods. For example,
\begin{Schunk}
\begin{Sinput}
R> plot(pl_m)
\end{Sinput}
\end{Schunk}
creates a log-log plot of the data, while the \cc{lines} function
\begin{Schunk}
\begin{Sinput}
R> lines(pl_m, col=2)
\end{Sinput}
\end{Schunk}
adds the fitted distribution (to get Figure~\ref{F1}a). When calling the
\texttt{plot} and \texttt{lines} function, the data plotted is 
invisibly returned, i.e.
\begin{Schunk}
\begin{Sinput}
R> dd = plot(pl_m)
R> head(dd, 3)
\end{Sinput}
\begin{Soutput}
  x      y
1 1 1.0000
2 2 0.5141
3 3 0.3505
\end{Soutput}
\end{Schunk}
This makes it straightforward to create graphics using other R packages.

To fit other distributions, we follow a similar procedure. For example, to fit
the discrete log normal distribution, we begin by creating a \cc{dislnorm}
object and estimating the parameters\footnote{\cc{dislnorm}: \textbf{dis}crete
  \textbf{l}og-\textbf{norm}al.}
\begin{Schunk}
\begin{Sinput}
R> ln_m = dislnorm$new(moby)
R> (est_ln = estimate_xmin(ln_m))
\end{Sinput}
\end{Schunk}
Then we update the object
\begin{Schunk}
\begin{Sinput}
R> ln_m$setXmin(est_ln)
\end{Sinput}
\end{Schunk}
and add the corresponding line to the plot
\begin{Schunk}
\begin{Sinput}
R> lines(ln_m, col=2)
\end{Sinput}
\end{Schunk}
giving Figure~\ref{F1}a. 

Figure~\ref{F1} gives example data sets, with associated power law and log
normal fits. Plotting the data in this manner has two clear benefits. First, it
highlights how much data is being discarded when fitting $\xmin$. Second,
it provides an easy comparison with other distributions.

\subsection{Parameter uncertainty}

\begin{figure}[t]
 \centering
\includegraphics[width=0.7\textwidth]{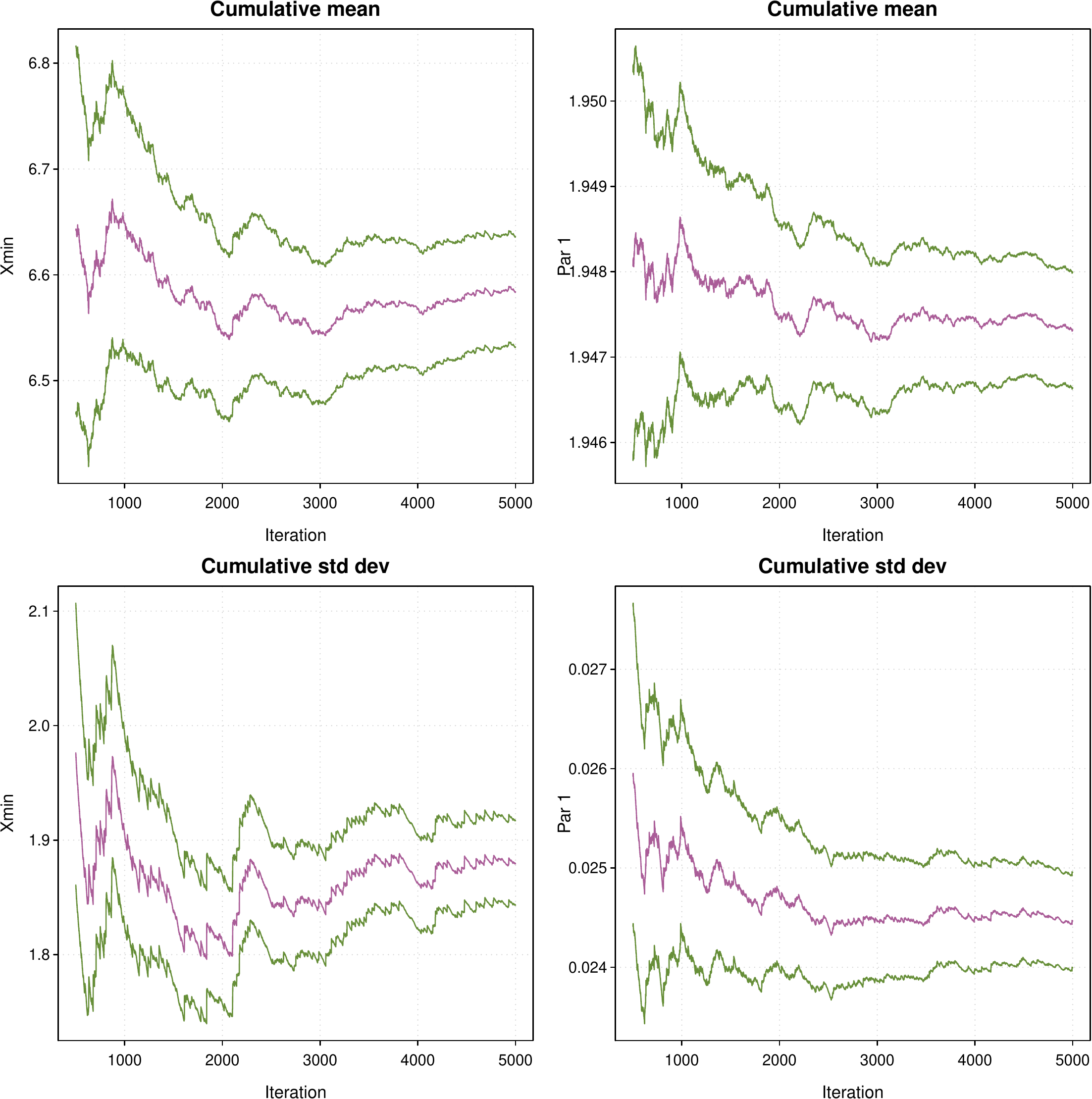}
\caption{Results from the standard bootstrap procedure (for the power law model)
  using the Moby Dick data set: \mbox{\texttt{bootstrap(pl\_m)}}. The top row
  shows the sequential mean estimate of parameters $\xmin$ and $\alpha$. The
  bottom row shows the sequential estimate of standard deviation for each
  parameter. The dashed-red lines give approximate 95\% confidence intervals. After
  5000 iterations, the standard deviation of $\xmin$ and $\alpha$ is estimated
  to be 1.9 and 0.02 respectively.}\label{F2}
 \end{figure}                    %

To get a handle on the uncertainty in the parameter estimates, we use a
bootstrapping procedure, via the \texttt{bootstrap} function. This procedure can
be applied to \textbf{any} distribution object. Furthermore, the bootstrap
procedure can utilize multiple CPU cores to speed up inference using the
built-in \pkg{parallel} package \citep{RCoreTeam2013}. To generate five thousand
bootstrap samples, using two cores, we use the following command
\begin{Schunk}
\begin{Sinput}
R> bs = bootstrap(pl_m, no_of_sims=5000, threads=2)
\end{Sinput}
\end{Schunk}
By default the \cc{bootstrap} function will use the maximum likelihood
estimate to infer the parameter values and check all values of $\xmin$. When
the $\xmin$ search space is large, then it is recommend that it is truncated.
For example
\begin{Schunk}
\begin{Sinput}
R> bootstrap(pl_m, xmins = seq(2, 20, 2))
\end{Sinput}
\end{Schunk}
will only calculate the Kolmogorov-Smirnoff statistics at values of $\xmin$ equal to
\[
2, 4, 6, \ldots, 20\;.
\]
A similar argument exists for the parameters.

The bootstrap function returns a \cc{bs\_xmin} object. This object is a list
that consists of three components:
\begin{enumerate}
\item \cc{gof}: the goodness of fit statistic obtained from the Kolmogorov-Smirnoff test. This value should correspond to the value obtained from the \mbox{\cc{estimate\_xmin}} function;
\item \cc{bootstraps}: a data frame containing the results from the bootstrap procedure;
\item \cc{sim\_time}: the average simulation time, in seconds, for a single bootstrap.
\end{enumerate}
The bootstrap results can be explored in a variety way. First we can estimate the standard deviation of the parameter uncertainty, i.e.
\begin{Schunk}
\begin{Sinput}
R> sd(bs$bootstraps[,2])
\end{Sinput}
\begin{Soutput}
[1] 1.879
\end{Soutput}
\begin{Sinput}
R> sd(bs$bootstraps[,3])
\end{Sinput}
\begin{Soutput}
[1] 0.02447
\end{Soutput}
\end{Schunk}
Alternatively, we can visualise the results using the \cc{plot} method
\begin{Schunk}
\begin{Sinput}
R> plot(bs, trim=0.1)
\end{Sinput}
\end{Schunk}
to obtain Figure~\ref{F2}. The top row of graphics in Figure~\ref{F2} give a
sequential 95\% confidence interval for mean estimate of the parameters. The
bottom row of graphics give a 95\% confidence interval for the standard
deviation of the parameters. The parameter \texttt{trim} in the \texttt{plot}
function controls the percentage of samples displayed. When \cc{trim=0}, all
iterations are displayed. When \texttt{trim=0.1}, we only display the final 90\%
of data.
\begin{figure}[t]
\centering 
\includegraphics[width=\textwidth]{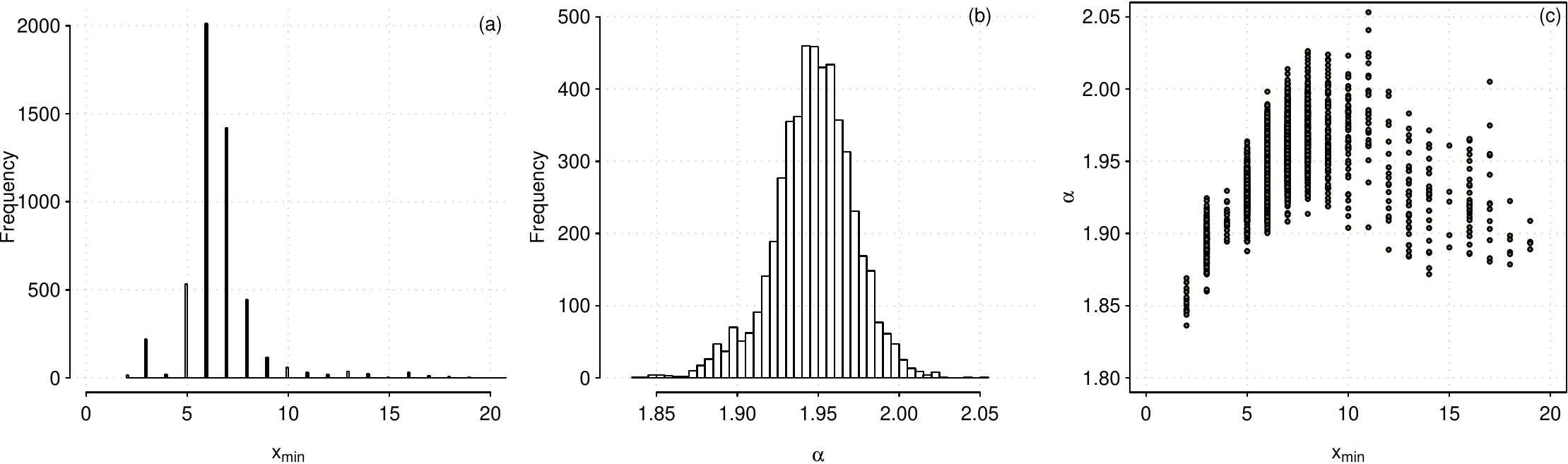}
\caption{Characterising uncertainty in parameter values using five thousand
  bootstraps. (a) Histogram of $\xmin$ (standard deviation 1.9). (b) Histogram of
  $\alpha$ (std dev. 0.02). (c) Scatter-plot of the $\xmin$ against $\alpha$.
}\label{F3}
\end{figure}

We can also construct histograms of the parameters
\begin{Schunk}
\begin{Sinput}
R> hist(bs$bootstraps[,2], breaks="fd")
R> hist(bs$bootstraps[,3], breaks="fd") 
\end{Sinput}
\end{Schunk}

to get Figures~\ref{F3}a \& b. A joint scatter plot is useful in highlighting
the strong dependency that often exists between the scaling parameter $\alpha$
and $\xmin$
\begin{Schunk}
\begin{Sinput}
R> plot(bs$bootstraps[,2], bs$bootstraps[,3])
\end{Sinput}
\end{Schunk}
and yields Figure~\ref{F3}c.

A similar bootstrap analysis can be obtained for the log normal distribution
\begin{Schunk}
\begin{Sinput}
R> bootstrap(ln_m)
\end{Sinput}
\end{Schunk}
In this case we would obtain uncertainty estimates for both of the log normal parameters.

\subsection{Comparison to other distributions}

The main thrust of \cite{Stumpf2012} is that many of the systems that are
characterised as having a power law distribution, could equally come from another
heavy tailed distribution. The \pkg{poweRlaw} package provides two methods for
testing the power law hypotheses.

\begin{figure}[t]
\centering 
\includegraphics[width=\textwidth]{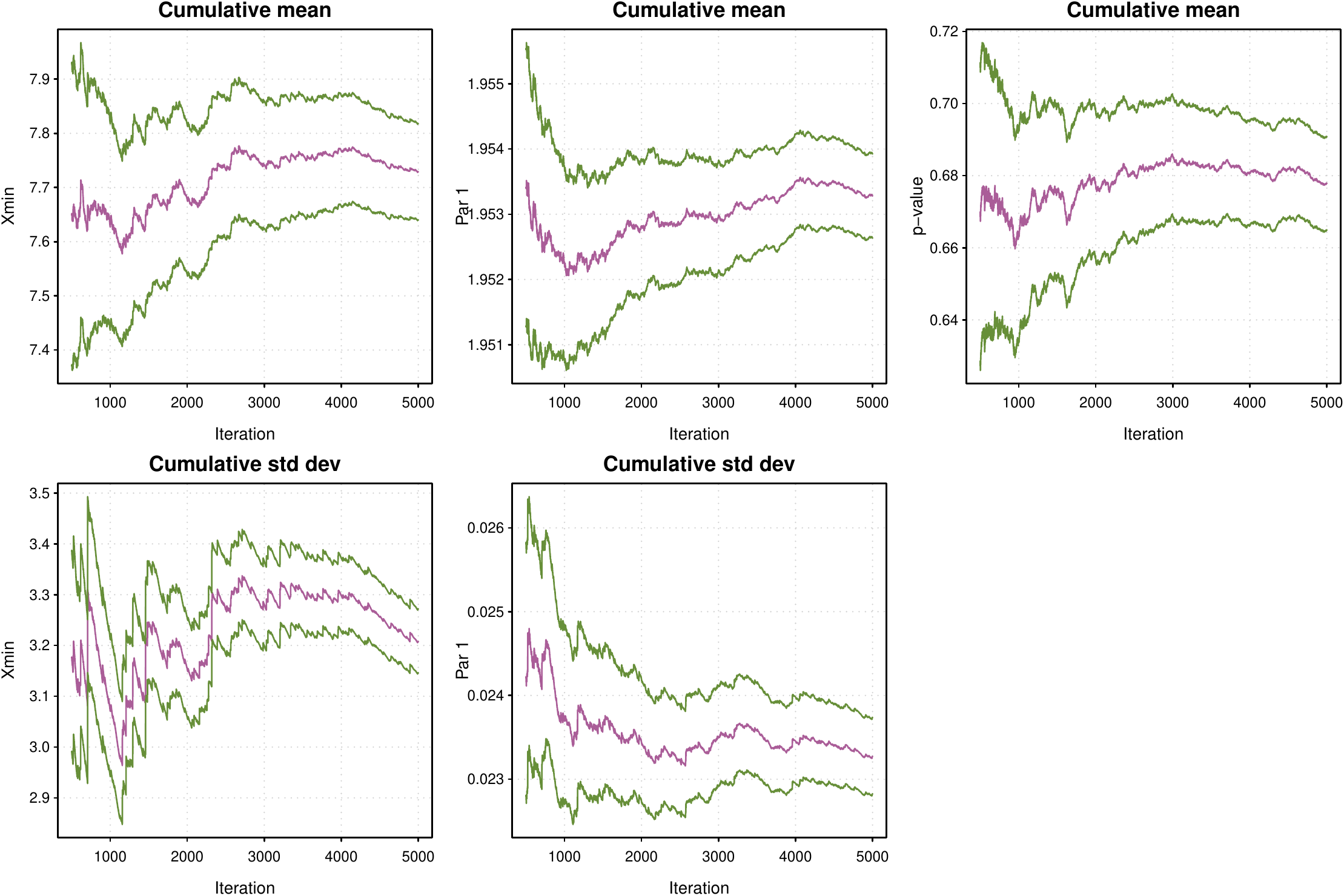}
\caption{Results from the bootstrap procedure for determining the plausibility
  of the power law hypothesis for the Moby Dick data set:
  \mbox{\texttt{bootstrap\_p(m\_pl)}}. The top row shows the sequential mean
  estimate of parameters $\xmin$, $\alpha$ and the $p\,$-value. The bottom row shows the
  sequential estimate of standard deviation for each parameter. The dashed-lines
  give approximate 95\% confidence intervals. }\label{F4}
\end{figure}

The first method uses the bootstrapping technique described in Algorithm~\ref{A2}. This
is accessed using a similar interface as the standard bootstrap function
\begin{Schunk}
\begin{Sinput}
R> bs_p = bootstrap_p(pl_m)
\end{Sinput}
\end{Schunk}
Again this function can be run in parallel (using the \cc{threads} argument) and
has the option to restrict the $\xmin$ search space. The output from the
\cc{bootstrap\_p} function has very similar structure to the \cc{bootstrap}
function. However, this function does return one additional element -- the
$p\,$-value for the hypothesis test:
\begin{align*}
&H_0: \text{the power law distribution can not be ruled out;}\\
&H_1: \text{the power law distribution can be ruled out.}
\end{align*}
In this particular example, we estimate $p=0.678$, i.e. the
underlying distribution for generating the Moby Dick data set could be a power
law. Again, the output can be easily visualised
\begin{Schunk}
\begin{Sinput}
R> plot(bs_p)
\end{Sinput}
\end{Schunk}

to obtain Figure~\ref{F4}. Notice that Figure~\ref{F4} has an additional plot
for the $p\,$-value. This enables the user to assess the accuracy of the estimated
$p\,$-value.

The second method is to directly compare two distributions using a likelihood
ratio test. For this test, both distributions must use the same $\xmin$ value.
For example, to compare the power law model to the log normal, we first the set
threshold to be the same as the power law model
\begin{Schunk}
\begin{Sinput}
R> ln_m = dislnorm$new(moby)
R> ln_m$setXmin(7)
\end{Sinput}
\end{Schunk}
Next we estimate the parameters (conditional on $\xmin=7$)
\begin{Schunk}
\begin{Sinput}
R> est = estimate_pars(ln_m) 
\end{Sinput}
\end{Schunk}
and update the model
\begin{Schunk}
\begin{Sinput}
R> ln_m$setPars(est)
\end{Sinput}
\end{Schunk}
Then we can use Vuong's method to compare models
\begin{Schunk}
\begin{Sinput}
R> comp = compare_distributions(pl_m, ln_m)
\end{Sinput}
\end{Schunk}

The object \cc{comp} object contains Vuong's test statistic, $p\,$-values and
the ratio of the log likelihoods. For this particular comparison, we have
$p=0.682$ which relates to the hypotheses
\begin{align*}
&H_0: \text{Both distributions are equally far from the true distribution;}\\
&H_1: \text{One of the test distributions is closer to the true distribution.}
\end{align*}
Hence, we can not reject $H_0$ and it isn't possible to determine which is the
best fitting model.

\section{Package overview }

In the previous example, we created a \texttt{displ} object
\begin{Schunk}
\begin{Sinput}
R> pl_m = displ$new(moby)
\end{Sinput}
\end{Schunk}
to represent the discrete power law distribution. This particular object has
class \cc{displ} and also inherits the \texttt{discrete\_distribution} class.
Other available distributions are given in Table~\ref{T2}. 
\begin{table}[t]
  \centering
  \begin{tabular}{@{} lll @{}}
    \toprule
    Distribution & Class name & \# Parameters \\
    \midrule
    Discrete power law & \texttt{displ} & 1 \\
    Discrete log normal & \texttt{dislnorm} & 2 \\
    Discrete exponential & \texttt{disexp} & 1 \\
    Poisson & \texttt{dispois} & 1 \\
    \\
    Continuous power law & \texttt{conpl} & 1 \\
    Continuous log normal & \texttt{conlnorm} & 2 \\
    Exponential & \texttt{conexp} & 1 \\
    \bottomrule
  \end{tabular}
  \caption{Available distributions in the \pkg{poweRlaw} package. Each class
    also inherits either the \cc{discrete\_distribution} or \cc{ctn\_distribution} class.}\label{T2}
\end{table}

\begin{figure}[t]
\centering 
\includegraphics[width=0.9\textwidth]{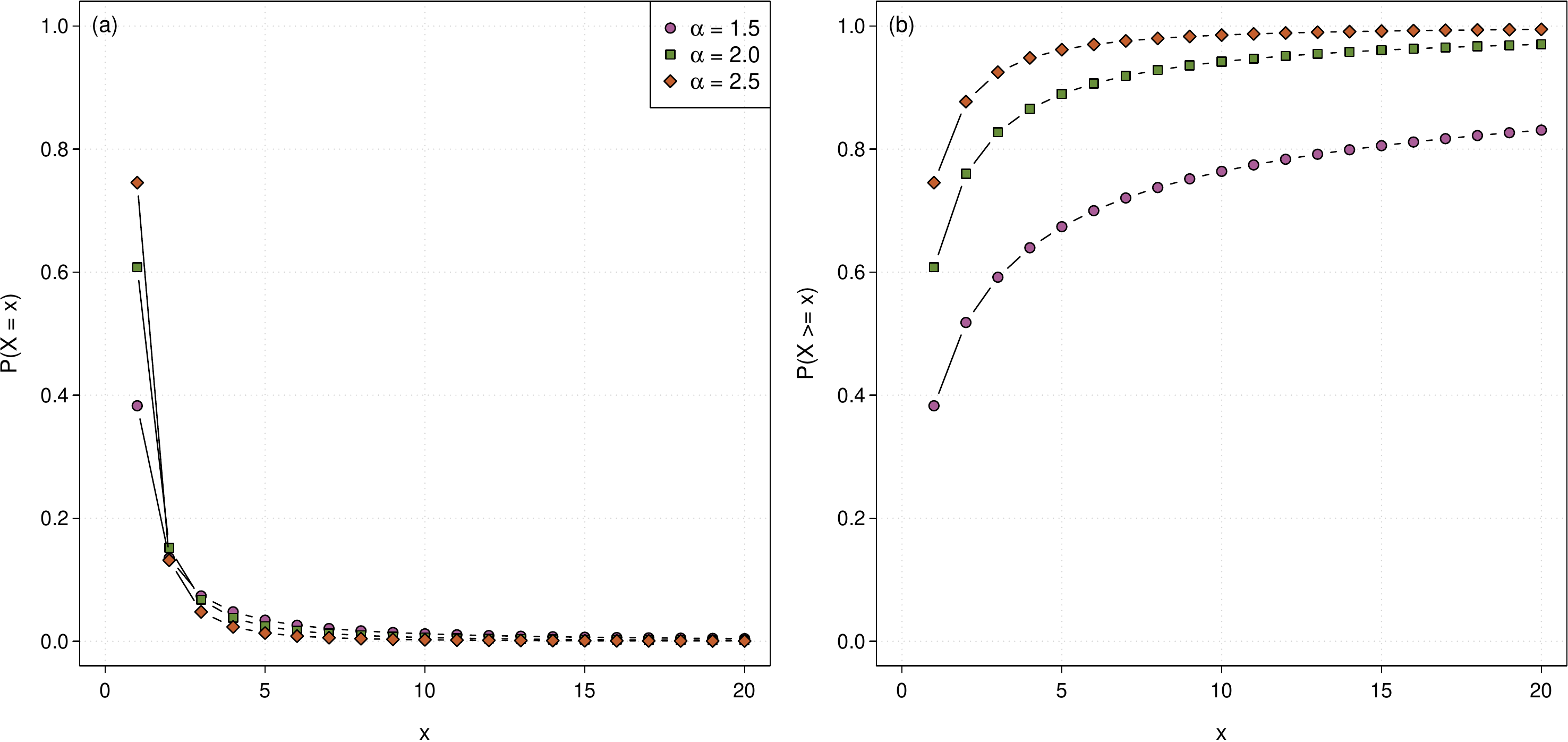}
\caption{The (a) probability mass function and (b) probability distribution
  function for the discrete power law, where $\xmin=1$ and $\alpha$ as indicated.}\label{F5}
\end{figure}

The classes given in Table~\ref{T2} are S4 reference classes\footnote{See
  \texttt{\mbox{?setRefClass}} for further details on references classes.}. Each
distribution object has four fields:
\begin{itemize}
\item \texttt{dat}: the data set;
\item \texttt{xmin}: the lower cut-off $\xmin$;
\item \texttt{pars}: a vector of parameter values;
\item \texttt{internal}: a list of values used in different numerical
  procedures. This will differ between distribution objects. In general, the
  user will not interact with the \cc{internal} field.
\end{itemize}
Using this particular object orientated framework has two distinct benefits. 
\begin{enumerate}
\item After fitting a single distribution, fitting all other distributions
  follows an almost identical route.
\item It is straightforward to add new distributions to the package.
\item The \cc{internal} field allows efficient caching of data structures when
  updating the \cc{xmin} and \cc{pars} fields. In particular, when the data is
  first loaded, efficient vector operations can be carried out and used as a
  look-up table, i.e. taking $\log$'s of the data.
\end{enumerate}
Distribution objects have a number of methods available (see Table~\ref{T3}).
All \cc{dist\_*} methods depend on the \textit{type} of distribution. For
example, to plot the probability mass function of the discrete power law
distribution, we first create a discrete power law object
\begin{Schunk}
\begin{Sinput}
R> m = displ$new()
R> m$setXmin(1)
\end{Sinput}
\end{Schunk}
then use the \cc{dist\_pdf} function to obtain the probabilities for particular
parameter values
\begin{Schunk}
\begin{Sinput}
R> x = 1:20
R> m$setPars(1.5)
R> plot(x, dist_pdf(m, x), type="b")
R> m$setPars(2.0)
R> lines(x, dist_pdf(m, x), type="b")
R> m$setPars(2.5)
R> lines(x, dist_pdf(m, x), type="b")
\end{Sinput}
\end{Schunk}
This gives Figure~\ref{F5}a. Likewise, to obtain the cumulative distribution
function we use the \cc{dist\_cdf} function, i.e. 
\begin{Schunk}
\begin{Sinput}
R> plot(x, dist_cdf(m, x), type="b")
\end{Sinput}
\end{Schunk}
to obtain Figure~\ref{F5}b.

The other methods, \cc{estimate\_*} and \cc{bootstrap\_*}, work with general
distribution objects (although internally they use \cc{dist\_*} methods). See
the associated help files for further details.
\begin{table}[t]
  \centering
  \begin{tabular}{@{} lp{10cm} @{}}
    \toprule
    Method Name & Description \\
    \midrule
    \texttt{dist\_cdf} & Cumulative density/mass function (CDF)\\
    \texttt{dist\_pdf} & Probability density/mass function (PDF)\\
    \texttt{dist\_rand}& Random number generator\\
    \texttt{dist\_data\_cdf} & Data CDF \\
    \texttt{dist\_ll} & Log likelihood\\
    \\
    \texttt{estimate\_xmin} & Point estimates of the cut-off point and parameter values\\
    \texttt{estimate\_pars} & Point estimates of the parameters (conditional on the current $\xmin$ value)\\
    \\
    \texttt{bootstrap} & Bootstrap procedure (uncertainty in $\xmin$)\\
    \texttt{bootstrap\_p} & Bootstrap procedure to test whether we have a power law\\
    \bottomrule
  \end{tabular}
  \caption{A list of methods available for \texttt{distribution} objects. These methods do not change the object states.}\label{T3}
\end{table}

\section{Conclusion}

In recent years an over-enthusiastic fitting of power laws to a wide variety of
systems has resulted in the inevitable (and needed) call for caution.
\cite{Stumpf2012} correctly highlight that many supposed power law relationships
are at best dubious and some obviously false. The error in determining the
underlying distribution of these mechanisms can (in some part) be placed at the
lack of available and easy to use software packages for fitting heavy tailed
distributions. The \pkg{poweRlaw} aims to solve this problem. By providing an
easy to use and consistent interface, researchers can now fit, and more
importantly, compare a variety of truncated distributions to their data set.

\bibliography{jss1180}

\end{document}